\documentclass[prb,showpacs]{revtex4}

\usepackage{amsmath,amssymb}

\usepackage{graphicx}

\usepackage{subfig}
\makeatletter
\newcommand{\justified}{%
  \rightskip\z@skip%
  \leftskip\z@skip}
\makeatother

\newcommand{\be}{\begin{eqnarray}} 
\newcommand{\ee}{\end{eqnarray}}

\usepackage{hyperref}

\usepackage{graphicx}

\begin{document}

\widetext

\title{Effect of the Pauli exclusion principle on the singlet exciton yield in conjugated polymers
}

\author{A. Thilagam}
\email{thilaphys@gmail.com}

\affiliation{Future Industries Institute, \\ 
University of South Australia,  Adelaide 5095, South Australia}
\begin{abstract}
Optical devices fabricated using conjugated polymer systems give rise to singlet exciton yields which
are high compared to
the statistically predicted estimate of 25\%  obtained using simple recombination schemes.
 In this study we evaluate  the singlet  exciton yield in conjugated polymers systems 
by  fitting to a model that incorporates the Pauli exclusion principle. 
The rate equations which describe the exciton dynamics include  quantum dynamical   components (both density and spin-dependent)
which arise during the spin-allowed
conversion of composite intra-molecular excitons into
loosely bound charge-transfer  (CT) electron-hole pairs.
Accordingly, a  crucial  mechanism by which singlet
excitons are increased at the expense of  triplet excitons is incorporated
in this work.  Non-ideal triplet excitons  which form at high densities, are rerouted via the
Pauli exclusion mechanism to form loosely bound CT states
which subsequently convert to singlet excitons.
Our derived expression for the yield in singlet exciton 
incorporates the  purity measure, and
provides a realistic description of the carrier dynamics  at   high exciton densities.
\end{abstract}
 
\pacs{78.67.-n, 77.22.-d}

\maketitle

\section{Introduction}
The properties of  singlet and triplet excitons (correlated electron-hole quasiparticles) in organic systems
 \cite{baldo2000transient,wilson2001spin,wohlgenannt2001formation,kersting,nguyen2000,friend1999electroluminescence,burroughes1990light,barford2004}  
 play  primary roles in determining
 the electroluminescent quantum efficiencies
of organic optical devices  fabricated using $\pi$-conjugated polymers 
\cite{skotheim2006conjugated,jenekhe1994excimers,shuai2000singlet,burroughes1990light,gunes2007conjugated,friend1999electroluminescence,sirringhaus,liu2014design}.
 Two fermions of spin $\frac{1}{2}$,  give rise to
 four microstates: one singlet state (antisymmetric under
particle exchange) and three triplet states (symmetric under particle exchange).
Unlike the triplet exciton, the singlet exciton undergoes rapid
radiative recombination  producing light. Accordingly,
the ratio of singlet to triplet exciton formation rates, $\gamma$, 
 provides a useful quantitative  measure of the  electroluminescence efficiency
in conjugated polymers \cite{wohlgenannt2001for,yao2014progress}.
Based on the assumption  that the formation cross
sections of singlet and triplet excitons  are the same,
$\frac{1}{4}$ of available bi-fermion  pairs (or charge carriers)
will form singlet excitons with the well documented statistical yield of 
 $\frac{1}{1+ 3}$ ( 25\%).
However several works \cite{shuai2013,carvelli2011} have shown  evidence of  internal
quantum efficiencies of 25\% or more. 
To this end, the exact processes of
 spin dynamics that determines  singlet exciton yield  and the 
 electroluminescence efficiencies are not fully resolved.

The conversion dynamics  between the singlet and triplet excitonic states, related intermediate
states which present as loosely bound electron-hole pairs, as well as the
 recombination lifetimes of all known states are fundamental to evaluating the yields
of singlet exciton states.  Several studies \cite{wohlgenannt2001formation,kim2000,lin2003triplet,barford2004} have examined
the underlying mechanisms for the high singlet exciton yields
($>$ 25\%) observed in experimental results involving conjugated polymer systems.
The decrease of a  spin-dependent $\gamma$ was associated with
 strong spin-orbital coupling  effects in an earlier work \cite{lin2003triplet}.
By taking into account the optical interference effects on radiative rates,
a higher singlet exciton yield (35\%-45\%)
was obtained in conjugated polymers \cite{kim2000}.
The singlet exciton generation is lower in monomers compared
to its generation in polymers. This may be linked to the fact that
 recombination dynamics is spin-independent
in monomers,  while it becomes  spin-dependent
in  polymers \cite{wilson2001spin}.
These differences highlight the dependence of exchange interactions and other electronic
properties  on the molecular configuration of the polymer chains \cite{schwartz2003}.
The  energy relaxation process involving the  singlet exciton 
is rapid unlike the slower relaxation via the low lying triplet exciton  \cite{kara2003spin}.
This factor  may contribute partly to differences in dynamics of the two spin states of the exciton.
It appears that several underlying processes  account for the
 variations in singlet exciton yield, which also dependes on the 
 material system under investigation. Thus far  the 
inclusion of the Pauli-exclusion principle which becomes dominant
at higher exciton densities has not been incorporated in  all
 earlier works \cite{wohlgenannt2001formation,barford2004,kim2000,lin2003triplet}. 
The importance of the Pauli-exclusion principle at increasing exciton
densities and its neglect in earlier works  forms the key motivation for the
current study.

While two or more  ideal excitons can occupy the same quantum state, the
multiple occupation of single-particle states is forbidden for 
several fermions by virtue of the  Pauli-blocking mechanism.  
The rapid absorption of photons under  intense
illumination conditions excites  multiple pairs of electrons and holes,
generating closely spaced excitons. The decrease in distance between the optically generated electron-pair pairs.
results in  non-ideal  bosonic excitons  that acquire fermionic attributes.
The ideal boson commutation relations which are applicable
to the structureless exciton  need to be modified to  non-ideal  commutation forms
due to the operation of  the Pauli exclusion principle \cite{law}. 
While the   exclusion principle  operates under all conditions,
 it does not influence the dynamics of almost ideal
 excitons at very low densities.

The quantum statistical attributes of the 
coboson such as  the composite exciton  \cite{comb,combescot2008many,combescot2009role} 
has received increased interest in  recent works \cite{law,woot,gav,kurzy,ticprl,ticbo,thil14}. These studies examined the links
between  entanglement and composite nature of bosons with non-ideal attributes.
A key result is that quantum entanglement  underpins the binding of   constituent fermions 
to form the  composite boson\cite{law,woot}. While large entanglement between constituent fermions gives rise
to a point-like structureless ideal boson, the less
entangled electron-hole pair represents the non-ideal composite boson.
Quantum entanglement of the fermions constituents is thus  critical to ensuring  the
bosonic attribute of the composite system  remains intact \cite{law,hash}. 

In this work, the concept of entanglement is incorporated via the purity measure $0 \le P \le 1$
to examine quantum processes in organic materials.  
The purity measure is linked to the number of Schmidt modes \cite{law,woot} associated with
the entanglement of the paired structure of the exciton or any other species of composite boson.
An exciton with zero or very small purity $\rightarrow$ 0 behaves as an ideal boson, while non-zero values of 
$P$ describe a non-ideal composite boson. 
A loosely bound electron-hole pair is  associated with
the less entangled electron and hole system with $P >$ 0. 
In a system of many excitons, the purity $P$ quantifies the correlations between the fermions, with
$P \approx $ 0 implicating a highly correlated  system of ideal  excitons.
For $N$ excitons that occupy the same quantum state,
the exciton can be considered a boson as long as $N P \ll 1$ \cite{law}.
Accordingly, $\frac{1}{P}$ can be interpreted as  the number of particles that can occupy the same pure state, before the 
Pauli blocking mechanism begins to influence  the idealized nature of the bosons.
 Multiple singlet and triplet excitons
 generated in organic molecules experience a crossover  to 
non-ideal bosonic states as the excitons are confined to spaces
which are   comparable to their size  \cite{thilcross,thilscat}.
In this study, we  employ a composite boson model of the  exciton
to examine the singlet exciton yield at non-zero values of the purity $P$.
The incorporation of entanglement aspects in this study  will help overcome
the shortcomings of conventional  models  used to model  correlated system behavior in solids.

We consider  the model of inter-chain electron-hole
interaction  between pairs of parallel polymers,
where  ``charge-transfer" (CT) exciton states exist
at intermediate energies levels positioned between 
the electron-hole continuum and the  strongly bound exciton
states  \cite{barford2004}.  The singlet and triplet charge-transfer (CT)  quasi-particles
appear as critical precursors to the tightly bound exciton states
in polymers. We extend this model by incorporating a mechanism
by which tightly bound excitons convert  to the more loosely
bound CT intermediate states with loss of pure bosonic attributes.
The conversion occurs due to operation of the Pauli exclusion principle,
with the charge-transfer states  possessing a higher
degree of fermionic attributes. 

The energetic difference between singlet and triplet exciton
which arise from the opposite sign of the  exchange interaction
is generally high ( $\approx$ 0.7 eV)  in conjugated polymers \cite{kohler2004singlet}.
The ``charge-transfer" (CT) singlet and triplet 
states, on the other hand are formed at approximately the same energy levels \cite{barford2004}
due to the weak exchange interactions of these states.
Inter-conversion between the two types (singlet, triplet) of intermediate CT states may occur
soon after their formation. 
Typical  exciton lifetimes for organic materials
such as poly(p-phenylene vinylene), PPV \cite{blom} and anthracene \cite{grover}
are of the order of 1 ns \cite{friend}.
In other organic materials such as PDHFV, PTEH and PDHFHPPV \cite{wu}, the exciton 
lifetimes can range from 0.05 to 0.71 ns. 

In this work, we consider that the Pauli exclusion 
 operational times may
proceed faster  than the exciton lifetimes at high densities.
The Pauli exclusion related conversion of one excitonic species to another
becomes favorable over excitonic interactions with phonons or impurities with
increased deviations in the bosonic attributes of excitons.
Depending on the purity, $P$ of the excitons and  
the  material system  \cite{friend,wu} under study,  
Pauli exclusion based conversions are expected to proceed at times in the
order of 0.01 to 1 ns.  These processes are expectedly absent at low  densities 
due to the ideal bosonic nature of the excitons.
The results in this study are thus relevant to excitons which become non-ideal bosons
as the excitonic  purity is increased.
In this study, we provide results of the singlet exciton yield for  range of 
conversion times linked to the Pauli mechanism. The main results  of this work
is  the derivation of an expression  for the singlet exciton yield
which incorporates the  purity measure.

\section{Rate equations incorporating the Pauli exclusion principle}

\begin{figure}[tbp]
\begin{center}
\includegraphics[scale=0.30]{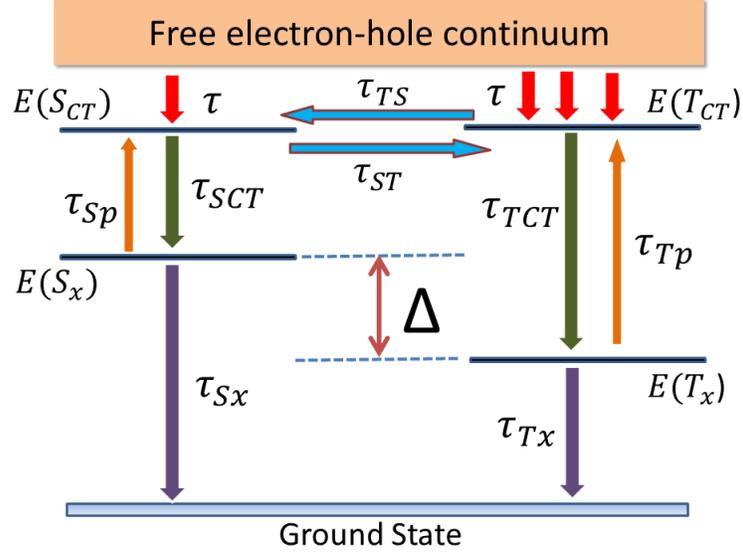}
\end{center}
\caption{
Schematics of the energy levels ($E(i)$) of  singlet ($S_X$) and triple ($T_X$) exciton states 
and  singlet ($S_{CT}$) and triplet ($T_{CT}$)
``charge-transfer" states. The  conversion times  within the
singlet states ($\tau_{SCT}$) and triplet states ($\tau_{TCT}$) 
are indicated by the green arrows. The unequal times of inter-conversions between the spin CT states
  ($\tau_{TS}$, $\tau_{ST}$)  are indicated by the blue arrows. 
The  Pauli exclusion  reversal times ($\tau_{Tp}$ and $\tau_{Sp}$), which are present only
at non-zero exciton purity $p$, are denoted by the orange arrows,
while the  recombination times of the singlet and triplet excitons ($\tau_{Sx}$, $\tau_{Tx}$)
are shown via the purple arrows.  $\Delta$ denotes
the  exchange energy between $S_X$ and $T_X$. } 
\label{mod}
\end{figure}

In typical organic devices, the injection of  charges carriers  precedes the formation of
excitons which  diffuse for an amount of time before  decaying radiatively (singlet exciton) or non-radiatively (triplet exciton). 
The overall external quantum efficiency of the organic devices is written as a product, $\eta_f \; \eta_s \; \eta_p \; \eta_o$
where $\eta_f$ is the efficiency of exciton formation
from free carriers, $\eta_s$ is the singlet exciton yield, $\eta_p$ is the 
photoluminescence quantum efficiency  and $\eta_o$ is the fraction of 
of photons which are emitted without being utilized by the device.
When the exciton density becomes high enough, the manifestation of the
Pauli exclusion principle can be examined through    $\eta_f$ and $\eta_o$,
 with incorporation  of relevant  device parameters.
In this work, we focus  on the effects of the Pauli exclusion principle on the underlying processes which
determine  $\eta_s$. To evaluate  $\eta_s$, we consider a typical
setup in which both singlet and triplet excitons are involved
in a coordinated process to produce the optimal
singlet exciton yield.

The schematics of the energy levels of  singlet and triple exciton states and ``charge-transfer" (CT)
states for the model under study is shown in Fig.\ref{mod}.
We  employ several notations adopted in Ref. \cite{barford2004}
for exciton and CT states, and assume that charge carriers
which are injected into the polymer material possess
random spin orientations.  The injected charge carriers 
 form the loosely bound charge-transfer singlet and triplet quasi-particles,
$S_{CT}$ and $T_{CT}$. The   inter-system-crossing  between
$T_{CT}$ and $S_{CT}$ due to the spin-orbit coupling mechanism
 is considered unequal, where
$\tau_{TS}$ ($\tau_{ST}$) denotes the time taken for
$T_{CT}$ ($S_{CT}$) state to convert to the $S_{CT}$ ($T_{CT}$) state.
The conversion from $T_{CT}$ to the triplet exciton, $T_X$ occurs
at the rate, $1/\tau_{TCT}$, while  the conversion from $S_{CT}$ to the triplet exciton, $S_X$ occurs
at the rate, $1/\tau_{SCT}$. 
The  recombination of the singlet (triplet) exciton state $S_X$ ($T_X$) at energy
level, $E(S_X)$ ($E(T_X)$) to the ground state occurs  in the time, $\tau_{Sx}$ ($\tau_{Tx}$).

We consider the occurrence of   Pauli exclusion reversal times  $\tau_{Tp}$ and $\tau_{Sp}$ 
 for the triplet and singlet exciton states respectively at non-zero exciton purity ($p_t, p_s$).
 Depending on the  degree of exciton purity or
compositeness, the reversal times  may proceed
faster  than the times of $T_r, \tau_{Sx}, \tau_{Tx}$.
The reversal times at the non-zero exciton purity $\tau_{Tp}$ and $\tau_{Sp}$
are indicated by the upward arrows in Fig.\ref{mod}.
Both these reversal times do not apply at low exciton densities,
for which  the  recombination processes, $S_{CT} \rightarrow S_X$ and $T_{CT} \rightarrow T_X$
occur  uninterrupted due to the  almost ideal bosonic exciton.
At high enough exciton densities where the purity assume non-trivial values,
Pauli exclusion related processes become favorable compared to
exciton decoherence due to interaction with phonons and/or impurities.
The role of the times, $\tau_{TS}$, $\tau_{ST}$, $\tau_{SCT}$, $\tau_{TCT}$, $\tau_{Tp}$ and $\tau_{Sp}$
in enhancing the singlet exciton yield will be examined
assuming  the absence of coherence between the singlet and triplet CT states.

We first consider the simplified model with 
equal times of inter-conversions between the spin CT states, so
that $\tau_{TS}$=$\tau_{ST}$=$\tau_{ISC}$.
The set of coupled equations relating the number of various
quasiparticle states such as
singlet excitons  ($N_{SX}$), triplet excitons ($N_{TX}$), singlet CT states
($N_{SCT}$) and triplet CT state ($N_{TCT}$) appear as
\be
\label{d1}
    \frac{d N_{SCT}} {dt} &=& \frac{N_i}{4\tau} +
    \frac{N_{TCT}} {\tau_{ISC}} - N_{SCT} \left( \frac{1}{\tau_{ISC}}
    + \frac{1}{\tau_{SCT}} \right)+ \frac{P_{s} N_{SX}} {\tau_{Sp}}, \\
\label{d2}
    \frac{d N_{TCT}} {dt} &=& \frac{ N_i}{2\tau} +
    \frac{N_{SCT}} {\tau_{ISC}} - N_{TCT} \left( \frac{1}{\tau_{ISC}}
    + \frac{1}{\tau_{TCT}} \right)+ \frac{P_{t} N_{TX}} {\tau_{Tp}}, \\
\label{d3}
    \frac{d N_{SX}} {dt} &=& \frac{N_{SCT}}{\tau_{SCT}} -
    \frac{N_{SX}} {\tau_{SX}}-\frac{P_s N_{SX}} {\tau_{Sp}}, \\
\label{d4}
    \frac{d N_{TX}} {dt} &=& \frac{N_{TCT}}{\tau_{TCT}} -
    \frac{N_{TX}} {\tau_{TX}}- \frac{P_t N_{TX}} {\tau_{Tp}}.
\ee
where $\frac{N_i}{\tau}$ is the initial number electron-hole pairs
created per second. To reduce the number of coupled equations,
we have used a single term ($N_{TCT}$) to denote the two possible triplet terms.
In Eqs. \ref{d1} to \ref{d4}, we have considered  an exciton number dependent term 
which  appears as 
\be
 \frac{p_{i} N_{jX}} {\tau_{jp}},
\label{foeq}
\ee
where $ N_{jX}$ denotes the number of
singlet exciton ($N_{SX}$) or  triplet excitons ($N_{TX}$). The 
 purity measure $P_{s}$ ($P_{t}$) is associated with the singlet (triplet) 
non-ideal exciton.  In order to obtain  quantitative results in forthcoming sections, we consider that the
 Pauli exclusion reversal times  $\tau_{Tp}$ and $\tau_{Sp}$ to be of the
same order as the recombination times of the singlet and triplet excitons $\tau_{Sx}$, $\tau_{Tx}$.
The form of  Eq. \ref{foeq} is based on the lower bound  of the normalization ratios
associated with $(N+1)$ and $N$ exciton which will be discussed next.

\subsection{Pauli exclusion purity factors, $P_s$ and $P_t$}

The  purity measures, $P_s$ and $P_t$ in Eqs. \ref{d1}-\ref{d4}  of the non-ideal singlet and triplet excitons  created 
from the ``charge-transfer" states  appear in the derived relation \cite{woot}
\be 
1- P_i\cdot N \le \frac{\chi_{N+1}}{\chi_N} \le 1-P_i,
\label{ineq}
 \ee
where $i=$ s (singlet) or t (triplet), and 
$\chi_N$ is the normalization term linked to  the superposed state of
  $N$ non-ideal excitons \cite{law,woot}. 
The deviations from unity in the  ratio, $ \frac{\chi_{N+1}}{\chi_N}$
yield a measure of the non-bosonic quality of the correlated electron-hole
pair \cite{comb,woot,law}, establishes a link with 
with entanglement.
The  upper and lower bounds
to $ \frac{\chi_{N+1}}{\chi_N}$ in Eq. \ref{ineq} 
appear in terms of the purity $P_i$ of the single-fermion reduced state.
The ratio  $ \frac{\chi_{N+1}}{\chi_N}$ can be interpreted as the 
effective probability of increasing the number of excitons
from  $N$ to $N+1$ in a system.  Excitons with ideal boson properties are described by
an $N$ independent term $\chi_N$ = 1 so that
$P_s$ = $P_t$ = 0.   The lower bound in Eq. \ref{ineq}
 decreases monotonically with $N$, and vanishes at $P$ = $\frac{1}{N}$.

The coherence between the $N$ excitons is decreased when  the $(N + 1)$st exciton
is added. This is due to the Pauli exclusion principle operating with
the probability, $1 - \frac{\chi_{N+1}}{\chi_N}$. 
The importance of incorporating density dependent
attributes, $P_i$ (i=s,t) in analyzing the singlet exciton yield
in Eqs. \ref{d1} to \ref{d4}  is highlighted by Eq. \ref{ineq}. 
The  density dependent
attributes, $P_i$ (i=s,t) therefore has to be taken into account in any analysis of the singlet or triplet 
exciton yield under strong illumination conditions.

At high purities, the  bounds in Eq. \ref{ineq} becomes less efficient
 \cite{ticprl,ticbo,tichy3} as the upper and lower bound terms
differ significantly from each other. The purity $P$ alone is not sufficient
to provide a reliable measure of the bounds in Eq. \ref{ineq}. In 
this regard $P$ cannot be used to accurately  predict  the bosonic attribute 
of excitons. To further exploit the simplicity of Eq. \ref{ineq},
we restrict our study to excitonic systems of small purity.
For the purpose of illustrating the
explicit role of the excitonic purity term, we  consider 
an $N$ independent upper bound for the  normalization ratio
\be
\frac{\chi_{N+1}}{\chi_N} \approx 1-P_i
\label{app2}
\ee
which is applicable to  a system of interacting excitons with minimal deviations from ideal
bosonic attributes. This simple model is used  to derive an analytical
form for the exciton yield. 
Subsequently, we utilize a realistic numerical model which is applicable
at small purity  based on  the relation
\be
P \approx \frac{\zeta}{N}
\label{app3}
\ee
 where  $0 < \zeta < 1$. 
We next employ Eqs. \ref{d1}-\ref{d4} as well as Eqs. \ref{app2} and \ref{app3} to obtain quantitative results of the exciton yield.

\subsection{Analytical form of the singlet exciton yield using Eq. \ref{app2}}

\begin{figure}[h]
\subfloat[]{\label{cap1}\includegraphics[width=6.5cm]{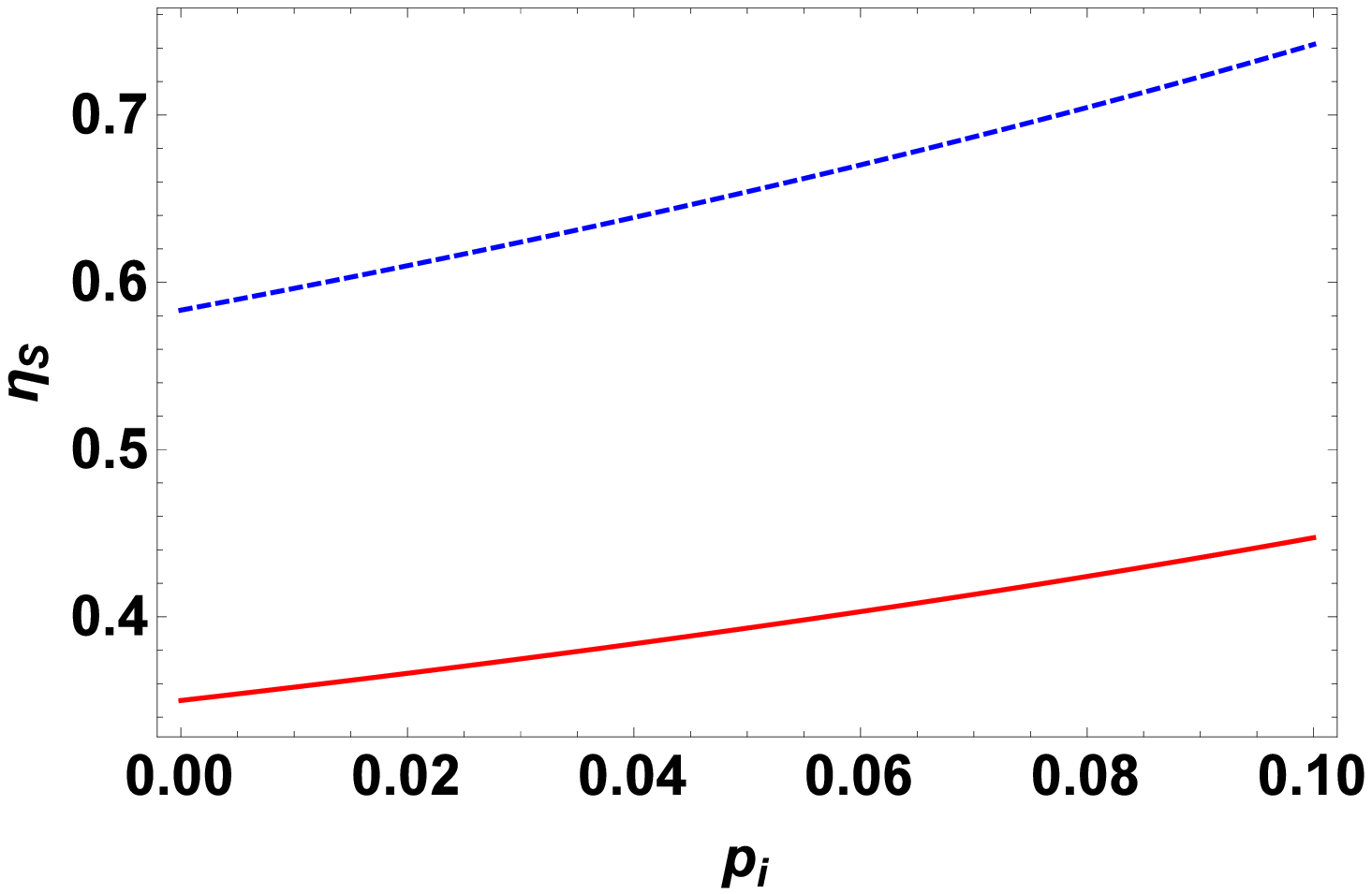}}\vspace{2.1mm} \hspace{2.1mm}
\subfloat[]{\label{cap2}\includegraphics[width=6.5cm]{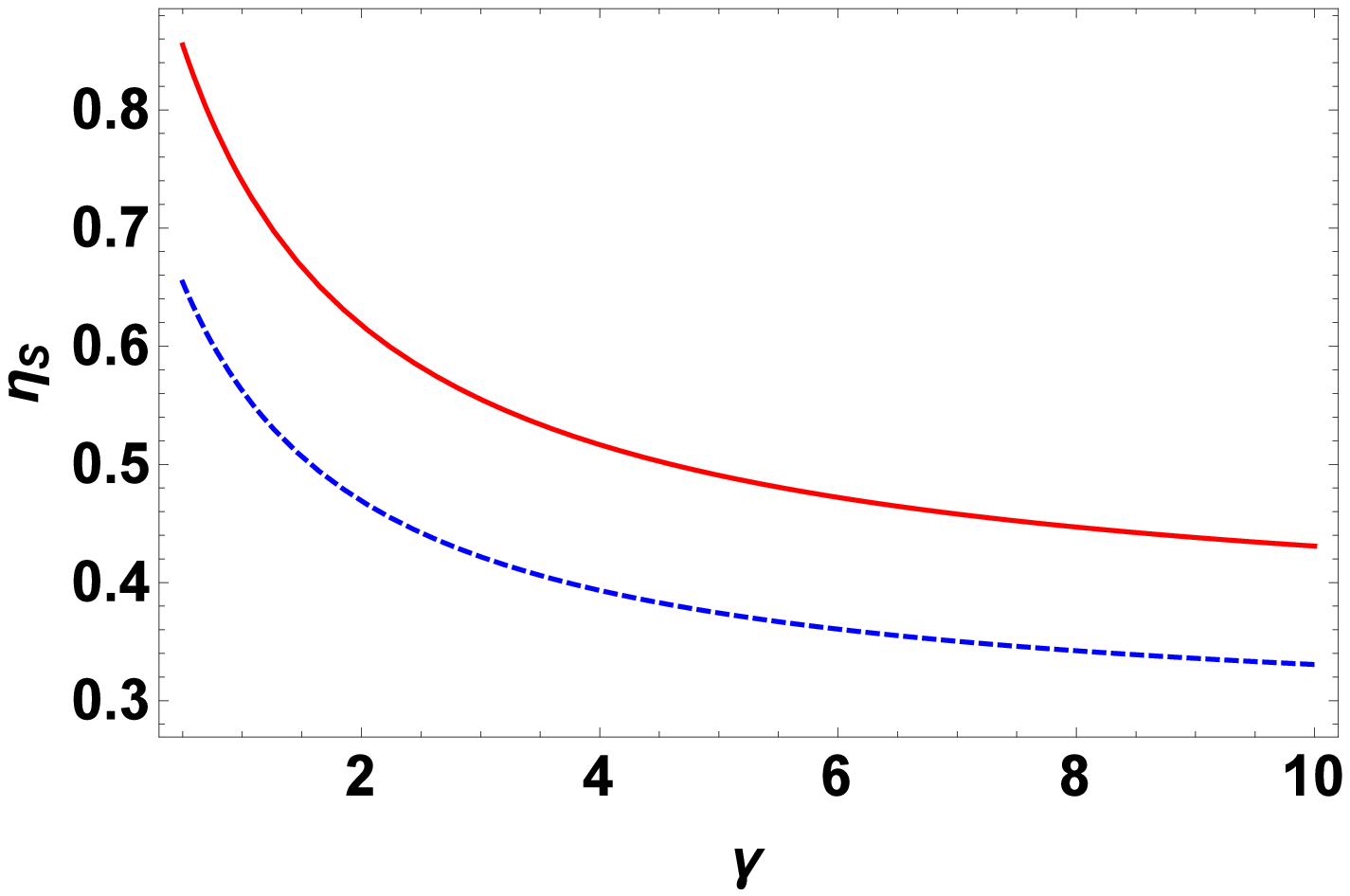}}\vspace{2.1mm} \hspace{2.1mm}
\caption{
(a) The singlet exciton yield, $\eta_s$ as a function of 
 $P_s$=$P_t$=$P_i$ obtained using Eq.\ref{ys}. The red solid line corresponds to
 $\gamma = \frac{\tau_{ISC}}{\tau_{TCT}}$= 4, while the blue dashed line corresponds to
 $\gamma $=0.5. \quad
(b) The singlet exciton yield, $\eta_s$ as a function of 
 $\gamma = \frac{\tau_{ISC}}{\tau_{TCT}}$ (Eq.\ref{ys}).
The red solid line corresponds to 
$P_i$=0.15, while the blue dashed line corresponds to
$P_i$=0.05. For both figures, we set $\alpha_s$=$\alpha_p$=1 
($\alpha_s =\frac{ \tau_{SX}}{\tau_{Sp}}$, $\alpha_p = \frac{\tau_{TX}}{\tau_{Tp}}$)
and $\beta$=0. }
 \label{yield}
\end{figure}

Using Eq.\ref{app2} which is applicable for system of weakly interacting bosons, and 
solving Eqs. \ref{d1} to \ref{d4} under the steady state conditions we obtain
 the singlet exciton yield $\eta_s$ as
\be
\nonumber
    \eta_s &=& \frac{N_{SX}/\tau_{SX}}{N_i/\tau} \\
&=&
\label{ys}
\frac{\gamma +3 P_t \alpha _t+3}
{4 \left(\beta +\gamma -\beta P_s \alpha _s-2 \gamma  P_s 
\alpha _s-2 P_s \alpha _s-2 P_s P_t \alpha _s \alpha _t+P_t \alpha _t+1\right)},
\ee
where  $\beta = \frac{\tau_{SCT}}{\tau_{TCT}}$ and $\gamma =
\frac{\tau_{ISC}}{\tau_{TCT}}$, $\alpha_s =\frac{ \tau_{SX}}{\tau_{Sp}}$ and
$\alpha_p = \frac{\tau_{TX}}{\tau_{Tp}}$.
 At  $P_s$ = $P_t$ = 0, Eq. \ref{ys}
reduces to the form
\be
\eta_s=\frac{3+\gamma}{4(1+\beta+\gamma)}
\label{app}
\ee
which was earlier  derived in  Ref.\cite{barford2004}. In
Fig. \ref{yield}a, the singlet yield is obtained as a  function of 
 $P_s$ = $P_t $= $P_i$ at $\beta$ = 0 (or $\tau_{TCT} \gg \tau_{SCT}$),
and for equivalent Pauli reversal times to recombination times, $\alpha_s$=$\alpha_p$=1.
Fig. \ref{yield}b shows the singlet yield as a  function of 
$\gamma = \frac{\tau_{ISC}}{\tau_{TCT}}$ at two different values of $P_s$=$P_t$=$P_i$.
We restrict the purity values to the range $0 \le P \le 0.15$  in Figs. \ref{yield}a, b 
as Eq. \ref{ys}  holds valid only for  small values of $P$.
 The results in Figs. \ref{yield}a, b  highlight the importance of
taking into account the Pauli exclusion mechanism  for non-ideal excitons. 
A gradual increase in the singlet exciton
yield occurs even with small increase in the purity values, $P_s$ and $P_t$.
This increase in the singlet exciton
yield is further enhanced at lower $\gamma$. These preliminary results suggest
that the spin-allowed route  $ T_X \rightarrow T_{CT}  \rightarrow  S_{CT} \rightarrow   S_{X}$
facilitates the increase in $\eta_s$. These findings can be compared with those of earlier works
\cite{kim2000,lin2003triplet,barford2004}
where an increase in the singlet exciton yield is attributed to transfer of excitation
from an intermediate triplet state rather than from the tightly bound 
triplet state.

The model used to 
derive results for Fig. \ref{yield}a,b is only applicable for small deviations
in bosonic attributes of the excitons. While we expect 
saturation in the singlet exciton yield to occur at higher $P$ resulting in $\eta_s \rightarrow 1$, this cannot be predicted
within the simple model used here. The usefulness of
Eq. \ref{ys} is highlighted by the fact that it  quantifies the mechanism by which there is increased proportion of  singlet
excitons through decrease in triplet exciton recombination.
  The triplet excitons are instead rerouted as a result of the
Pauli exclusion principle to form loosely bound CT states
which subsequently convert to singlet excitons.
As a consequence,  the fluorescence quantum yield 
can reach estimates of even 50\% or more, which is  the  main result of this work.

\subsection{Numerical solution for population of exciton species  using Eq. \ref{app3}}

 Using numerical techniques,  we  determine the population of
all species, $N_k$ ($k=SX, TX, TCT, SCT$) using Eqs. \ref{d1} to \ref{d4},
and  substituting $P_s$ and $P_t$ with the   $N$-dependent form 
in Eq.\ref{app3}.
We use typical parameters estimates for organic materials
such as poly(p-phenylene vinylene), PPV \cite{blom} and anthracene \cite{grover}
where the exciton lifetime is of the order of 1 ns \cite{friend},
$\tau_{TCT} \approx$ 10 ns \cite{barford2004}, 
$\tau_{SCT} \approx$ 1-10 ps \cite{grover} 
and $\tau_{ISC} \approx$ 0.1-10 ns \cite{kohler2004singlet,beljonne2001spin,grover}.

\begin{figure}[h]
\subfloat[]{\label{e1}\includegraphics[width=6.5cm]{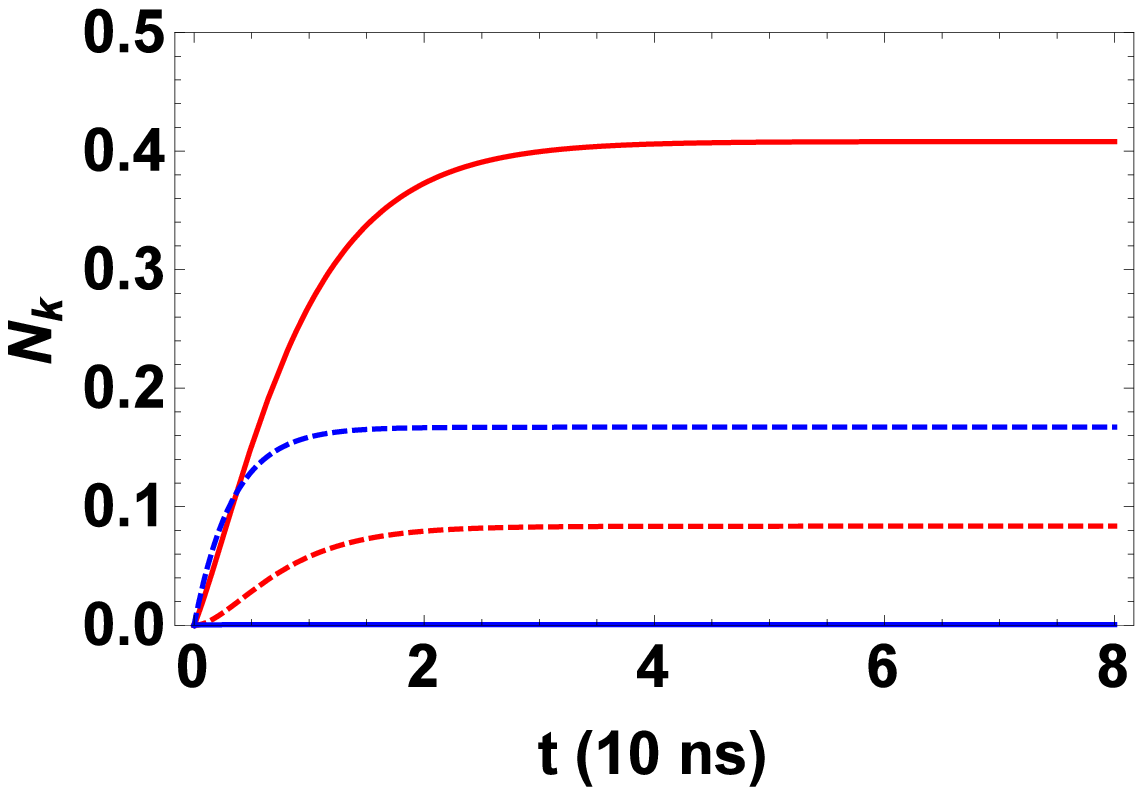}}\vspace{2.1mm} \hspace{2.1mm}
\subfloat[]{\label{e2}\includegraphics[width=6.5cm]{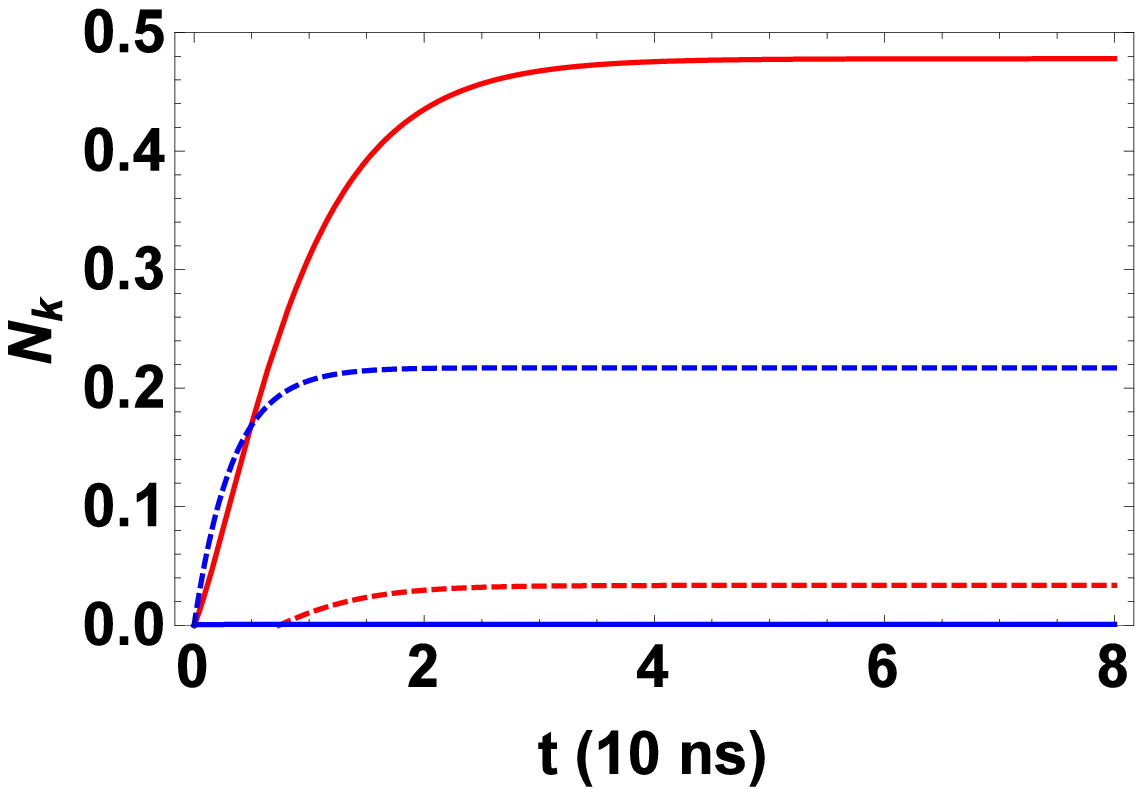}}\vspace{2.1mm} \hspace{2.1mm}
\caption{
(a)  $N_k$ (k=SX (red solid line), TX (red dashed line), SCT (blue solid line) , TCT (blue dashed line)) 
as a function of  time, $t$,
obtained by numerically solving  Eqs. \ref{d1} to \ref{d4}  using  Eq. \ref{app3}, and setting ${N_i}$=1,
${\tau}$=10 ns, $\zeta$ = 0.    We also choose $\beta$ = $\frac{\tau_{SCT}}{\tau_{TCT}}$=0.001,  $\gamma$ =
$\frac{\tau_{ISC}}{\tau_{TCT}}$=0.5, $\frac{\tau_{SX}}{\tau_{TCT}}$=0.7,
$\frac{\tau_{TX}}{\tau_{TCT}}$=0.5 with ${\tau_{TCT}}$= 10 ns.
\quad
(b) Same values for all parameters as in (a),  with exception 
of $\zeta$=0.15 (see Eq. \ref{app3}), and $\alpha_s$ =$\frac{ \tau_{SX}}{\tau_{Sp}}$=1, 
$\alpha_p$ =$\frac{\tau_{TX}}{\tau_{Tp}}$=1
}
 \label{evolve}
\end{figure}

The evolution of $N_k$ ($k=SX, TX, TCT, SCT$)
with time, $t$  is shown in Fig. \ref{evolve}.
While we set $\zeta$=0 in Fig. \ref{evolve}a,
a non-zero $\zeta$=0.15 (which reflects higher exciton purity)  is used to obtain  the results shown
 in  Fig. \ref{evolve}b. 
All other parameters are
provided in the caption of Fig. \ref{evolve}, in particular, we chose
${\tau_{TCT}}$= 10 ns and 
${\tau}$=10 ns. Accordingly the time of evolution, $t$ appears 
in units of 10 ns and  the population of
all species, $N_k$ are taken relative to the initial number electron-hole pairs,
 ${N_i}$ = 1. We note that for the chosen set of parameters,
the singlet exciton population rises rapidly and reaches a steady
state at $t \approx$ 30 ns.
 The triplet exciton which shows appreciable loss even at the beginning stage of evolution,
is  not apparent in Fig. \ref{evolve}a,b.
 The  population of the triplet CT species appears higher than the singlet CT species,
for the set of  parameters employed to solve Eqs. \ref{d1} to \ref{d4}. A  saturation 
of the single exciton population occurs beyond large times $t >$ 35 ns  for non-zero purity $P_s$=$P_t$.


\begin{figure}[h]
\subfloat[]{\label{e1}\includegraphics[width=6.5cm]{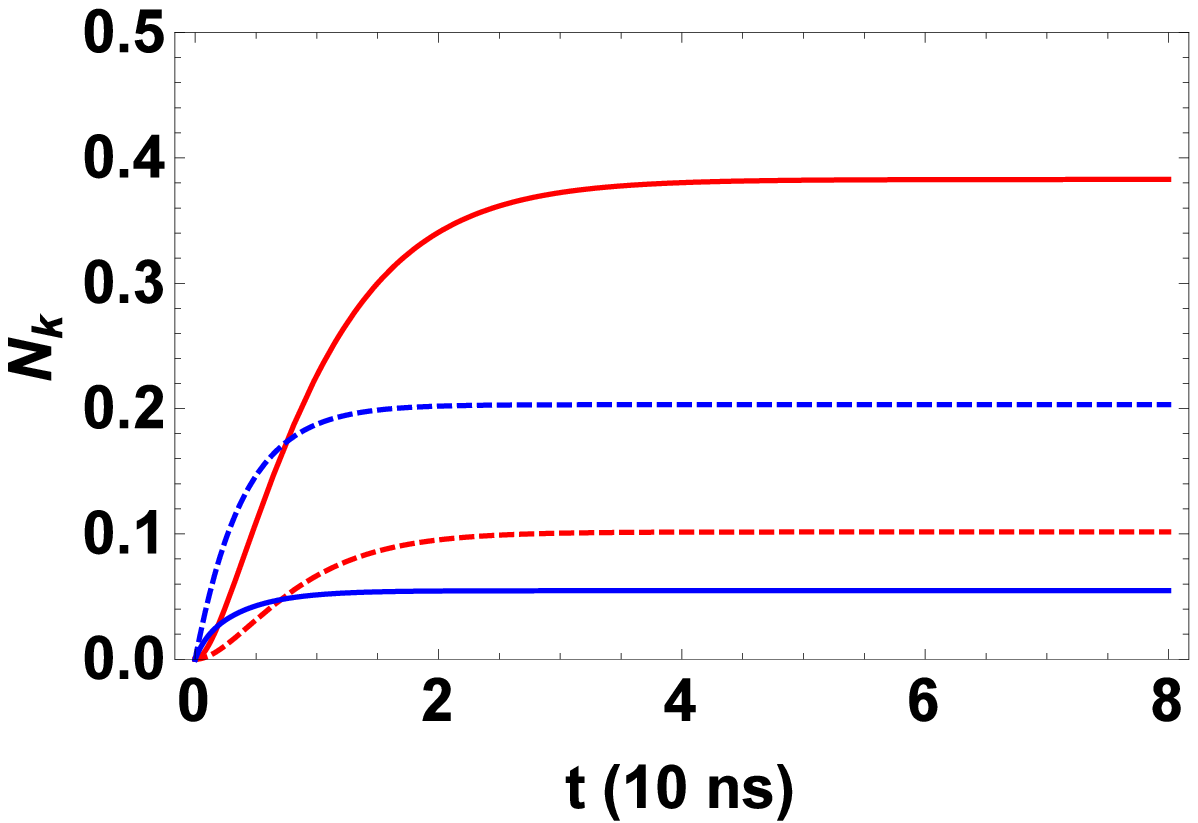}}\vspace{2.1mm} \hspace{2.1mm}
\subfloat[]{\label{e2}\includegraphics[width=6.5cm]{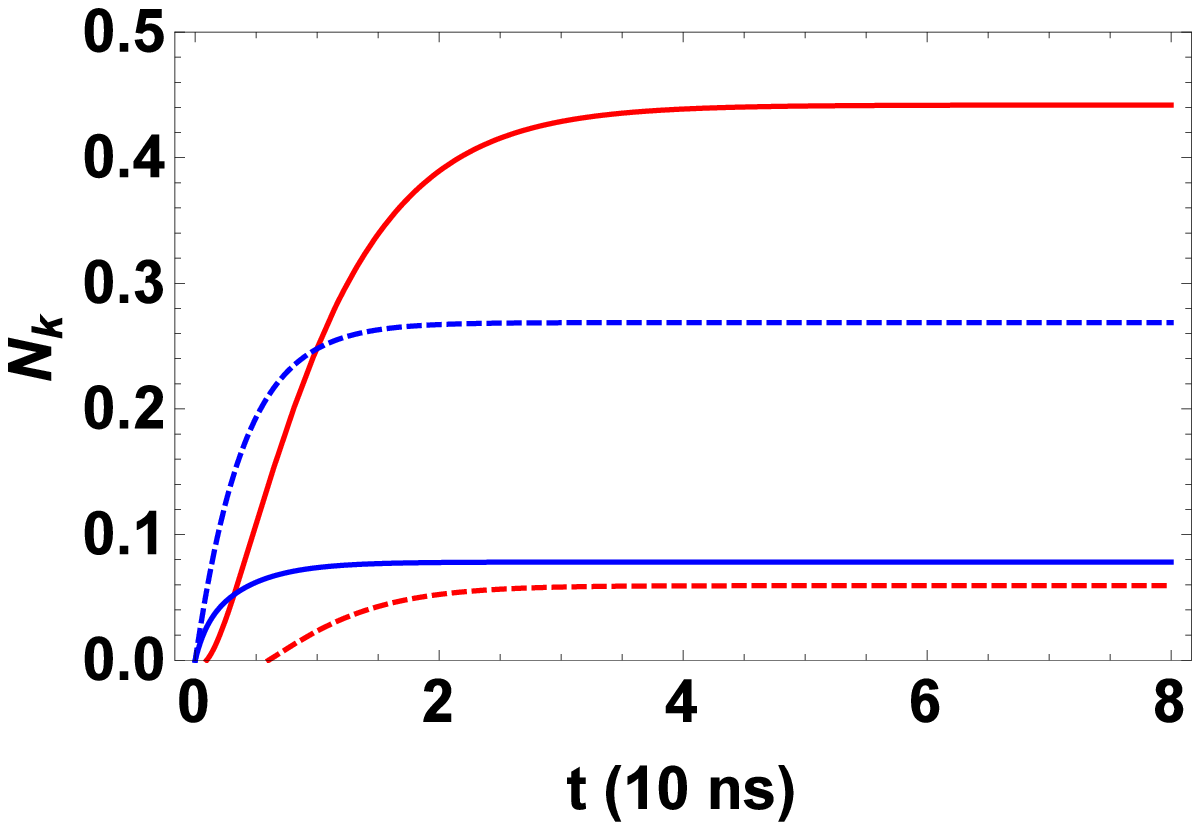}}\vspace{2.1mm} \hspace{2.1mm}
\caption{
(a)  $N_k$ (k=SX (red solid line), TX (red dashed line), SCT (blue solid line) , TCT (blue dashed line)) 
as a function of  time, $t$,
obtained by numerically solving  Eqs. \ref{d1} to \ref{d4} using  Eq. \ref{app3}, and setting ${N_i}$=1,
${\tau}$=10 ns, $\zeta$ = 0.    We set $\beta$ = $\frac{\tau_{SCT}}{\tau_{TCT}}$=0.1,  $\gamma$ =
$\frac{\tau_{ISC}}{\tau_{TCT}}$=0.5, $\frac{\tau_{SX}}{\tau_{TCT}}$=0.7,
$\frac{\tau_{TX}}{\tau_{TCT}}$=0.5 with ${\tau_{TCT}}$= 10 ns.
\quad
(b) Same values for all parameters as in (a),  with exception 
of $\zeta$=0.15, and $\alpha_s$ =$\frac{ \tau_{SX}}{\tau_{Sp}}$=1, 
$\alpha_p$ =$\frac{\tau_{TX}}{\tau_{Tp}}$=1
}
 \label{evolve2}
\end{figure}


In contrast to  Fig. \ref{evolve}a,b where we used $\beta$ = $\frac{\tau_{SCT}}{\tau_{TCT}}$=0.001,
the results in  Fig. \ref{evolve2}a,b are obtained using an increased  $\beta$ = 0.1.
Otherwise  other parameters used in Fig. \ref{evolve2}a,b remain the same as those used in 
Fig. \ref{evolve2}a,b. We note the prominent appearance of the triplet exciton 
with increase in $\beta$ in Fig. \ref{evolve2}a. Furthermore, with increase in the exciton purity
(non-zero $\zeta$), the population of the triplet exciton exceeds 
that of the singlet CT species. 
The results obtained here indicates the  possibility of  converting incoming photons 
to singlet excitons with minimal energy dissipation with onset of the Pauli exclusion 
mechanism under suitable conditions. The  results obtained in this study  clearly demonstrates
the critical role of triplet excitons when present at high densities.
We make clear that all results obtained in this study are valid for small exciton purity estimates.
The model employed in this study therefore stands valid for non-ideal excitons which depart slightly from possessing
ideal boson behavior.

\section{Conclusion}

In conclusion,  we  have 
derived an analytical relation of the singlet exciton yield that incorporates
the  Pauli exclusion mechanism via the purity measure, $P$.
Our model considers the  formation of singlet excitons via inter-conversion
from loosely bound  charge-transfer (CT) triplet electron-hole pair. We include a channel by which
 CT triplet states are  formed  from tightly bound excitons  at
high densities.  Hence we have  taken into account the crucial  mechanism by which singlet
excitons are increased by  preventing the loss of triplet exciton which otherwise would occur  through recombination
and other dissipation processes. Non-ideal triplet excitons  are rerouted via the
Pauli exclusion mechanism to form loosely bound CT states
which subsequently convert to singlet excitons.
As a consequence,  we obtain the fluorescence quantum yield 
that can increase beyond the well known estimate of  $> 25\%$. This is  the  main result of this work which 
 has importance   in  conjugated polymers systems.
The theoretical framework presented in this work provides  a possible explanation 
for the  high exciton yields
noted in earlier works \cite{wohlgenannt2001formation,kim2000,lin2003triplet}
 and   will be useful in evaluating optimal  electroluminescence efficiencies
in organic devices.

The idea of 
 entanglement control by structural variations
in novel quantum  dot nanostructures \cite{coe2010geometry,abdullah2009} 
may be employed to seek   enhancement of the singlet exciton
yield in polymer systems. 
The incorporation of the Pauli exclusion principle  during diffusion of
photogenerated excitons in
 drift$-$diffusion models \cite{buxton2007}  will also assist in the 
prediction of realistic
operating conditions in polymer based devices  with  morphologies
of different dimensions \cite{kodali2012,hoffmann,min2010optimization}. To this end,
future experimental investigations
may consider the control of  the entanglement attributes of
excitons through polymer length tailoring and engineering of
 morphological attributes \cite{ma2005thermally}.
The incorporation of the Pauli exclusion
principle in determining the efficiency of 
devices composed of other material systems such as CdSe and PbSe nanocrystals \cite{schaller2005high}
 where multiple excitons are generated, and layered transition  metal dichalcogenides
with enhanced excitonic  properties \cite{mak,rama,wilson,wang,thil2014}
may be explored in future works. Future experimental works
should seek to  investigate alternative means  of  targeting the
triplet excitons to increase the
singlet exciton yield.

\section{Acknowledgments}

This research was undertaken on the NCI National Facility in Canberra, Australia, which is
supported by the Australian Commonwealth Government.

\end{document}